\documentclass[english,conference]{IEEEtran}
\IEEEoverridecommandlockouts
\usepackage[T1]{fontenc}
\usepackage{amsthm}
\usepackage{amsmath}
\usepackage{mathrsfs}
\usepackage{amssymb}
\usepackage{graphicx}
\usepackage[latin9]{inputenc}

\makeatletter

\theoremstyle{plain}
\newtheorem{thm}{\protect\theoremname}
\theoremstyle{remark}
\newtheorem{rem}[thm]{\protect\remarkname}
\theoremstyle{plain}
\newtheorem{lem}[thm]{\protect\lemmaname}

\usepackage{babel}
\providecommand{\lemmaname}{Lemma}
\providecommand{\remarkname}{Remark}
\providecommand{\theoremname}{Theorem}

\newcommand{\cB}{\mathcal{B}}
\newcommand{\cP}{\mathcal{P}}
\newcommand{\cI}{\mathcal{I}}

\newcommand{\cK}{\mathcal{K}}

\makeatother

\begin{document}

\title{On the Duality of Fractional Repetition Codes}

\author{Bing Zhu$^{\dagger}$, Kenneth W. Shum$^{\ddagger}$, and Hui Li$^{\dagger}$\\
$^{\dagger}$School of Electronic and Computer Engineering, Peking University\\
$^{\ddagger}$Institute of Network Coding, The Chinese University of Hong Kong\\
E-mail: zhubing@sz.pku.edu.cn, wkshum@inc.cuhk.edu.hk, lih64@pkusz.edu.cn

\thanks{Hui Li is also with the Shenzhen Key Lab of Information Theory and Future Internet Architecture, Shenzhen Eng. Lab of Converged Networks Technology, and the Future Network PKU Lab of National Major Research Infrastructure, Shenzhen Graduate School, Peking University, China.}}

\maketitle

\begin{abstract}
Erasure codes have emerged as an efficient technology for providing data redundancy in distributed storage systems. However, it is a challenging task to repair the failed storage nodes in erasure-coded storage systems, which requires large quantities of network resources. In this paper, we study fractional repetition (FR) codes, which enable the minimal repair complexity and also minimum repair bandwidth during node repair. We focus on the \textit{duality} of FR codes, and investigate the relationship between the supported file size of an FR code and its dual code. Furthermore, we present a \textit{dual bound} on the supported file size of FR codes.
\end{abstract}


\section{Introduction}

In the recent decades, erasure coding techniques are gaining more and more popularity owing to the high storage efficiency. For the same storage overhead, erasure codes can improve the data reliability than the conventional replication-based scheme \cite{key-1}. Therefore, enterprise storage systems are also transitioning to erasure coding techniques for better storage features \cite{key-2,key-3}. In an erasure coding scheme, an original data object is mapped into $n$ coded blocks and spread across $n$ distinct storage nodes, satisfying that any $k$ out of the $n$ nodes are sufficient to retrieve the source file. Even though traditional erasure codes (such as Reed-Solomon codes) are capable of reducing the data storage overhead while preserving the fault tolerance guarantees, they are less efficient in failure recovery as compared to replication. When replicated data is lost, it can be simply reconstructed by downloading copies from other surviving nodes. However, for erasure codes, the conventional method for recovering the lost data is to reconstruct the original file by connecting to $k$ nodes and then re-encode to obtain the data stored in the failed node. In other words, repairing a single failed node requires the data from $k$ other nodes, resulting in a $k$ times higher reconstruction traffic than the replication scheme.

Regenerating codes are proposed in~\cite{key-4} aiming to reduce the amount of data transmission during the node repair operation. Specifically, an $(n,k,d,M,\alpha,\beta)$ regenerating code encodes a data object of size $M$ into $n\alpha$ coded packets, which are spread across $n$ storage nodes, with each node containing $\alpha$ packets. The stored data can be reconstructed by contacting any subset of $k$ nodes. Upon failure of a storage node, it can be replaced by a new node, whose content is generated by downloading $\beta$ packets from each of any set of $d$ surviving nodes. Particularly, for the case that the total amount of network transfer (i.e., $d\beta$) equals to the size of lost content stored in the failed node, then the corresponding code is referred to as a minimum-bandwidth regenerating (MBR) code~\cite{key-4}. However, the reduction in repair bandwidth is at the cost of imposing an encoding capability on the helper nodes, i.e., the $\beta$ transferred packets are obtained~as a function of the packets stored in the contacted node.

Rashmi~\textit{et al.}~\cite{key-5} introduced the notion of \textit{repair-by-transfer} for node repair, where each helper node transfers a portion of the stored data to the replacement node without any arithmetic operations. Such a repair model eliminates the computational burden at the helper nodes. El Rouayheb and Ramchandran~\cite{key-6} further generalized the code constructions in~\cite{key-5} and proposed fractional repetition codes. The encoding process of FR codes consists of the concatenation of two codes: an outer maximum-distance-separable (MDS) code followed by an inner fractional repetition code. The data objects are first encoded by an MDS code, and then the encoded packets are replicated and stored~in the system according to the FR code. The outer MDS code is used for data reconstruction, and the inner FR code is designed to facilitate the node recovery process. In the presence of node failures, the replacement node simply downloads the required data from a specific set of surviving nodes without the need~of computations. The transferred packets by the contacted helper nodes are the same as the content of the failed storage node. In such a sophisticated manner, FR codes achieve \textit{uncoded exact} repair at the MBR point.

El Rouayheb and Ramchandran presented in~\cite{key-6} two explicit constructions of FR codes based on regular graphs and Steiner systems. Several recent studies extend the design of FR codes to a larger set of parameters, which are based on randomized algorithm~\cite{key-7}, bipartite cage graphs~\cite{key-8}, resolvable designs~\cite{key-9}, incidence matrix~\cite{key-10}, group divisible designs~\cite{key-11}, transversal designs~\cite{key-12}, symmetric designs~\cite{key-13}, Kronecker product~\cite{key-14}, and perfect difference families~\cite{key-15}. Moreover, the supported file size of constructed FR codes are also investigated in \cite{key-12,key-14} and \cite{key-15}.

\textit{Contributions:} In this paper, we consider the duality of fractional repetition codes. Based on the relationship between~the supported file size of an FR code and its dual code, we present a dual bound on the supported file size of the original FR code.

\textit{Organization:} The rest of this paper is organized as follows. Section II introduces the duality of fractional repetition codes. We elaborate on the file size hierarchy of FR codes in Section III. Section IV presents a dual bound on the supported file~size of FR codes. Finally, we conclude in Section V.

\section{Fractional Repetition Code and Its Dual}

\begin{figure}
\centering{}\includegraphics[scale=0.115]{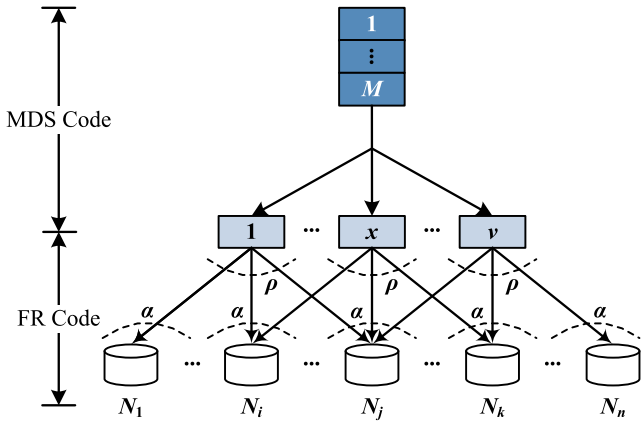}
\caption{The encoding process of an $(n,\alpha,v,\rho)$-FR code. A data object~of size $M$ is first encoded into $v$ coded packets using an MDS code. Then each coded packet is equally replicated $\rho$ times and distributed to $n$ storage nodes $N_1,N_2,\ldots,N_n$ according to the FR code, where each node stores $\alpha$ packets.}
\label{Encoding}
\end{figure}

An \textit{incidence structure} is a triple $(\cP,\cB,\cI)$, where $\cP$ and $\cB$ are two finite sets, and $\cI$ is a subset of $\cP\times\cB$. The elements in $\cP$ are called \textit{points}, and the elements in $\mathcal{B}$ are called \textit{blocks}. A point $p\in \cP$ is \textit{incident} with a block $B\in \cB$  if $(p,B)\in\cI$. The \textit{incidence matrix} of an incidence structure is a $|\cB|\times|\cP|$ zero-one matrix, with rows indexed by the blocks and columns indexed by the points, such that the entry corresponding to a point $p$ and a block $B$ is equal to $1$ if and only if $p$ is incident with $B$~\cite{key-16}.

In this setting, it is permissible that two distinct blocks are incident with the same set of points. We say that the~incidence structure contains \textit{repeated blocks}. For the case that there are no repeated blocks in an incidence structure $(\cP,\cB,\cI)$, we say that the incidence structure is \textit{simple}, and any block is~uniquely determined by the points incident with it. In other words, each block in $\cB$ can be viewed as a subset of $\cP$.

A \textit{fractional repetition (FR) code} is an incidence structure in which each point is incident with $\rho$ blocks, and each block is incident with $\alpha$ points, for some constants $\rho$ and $\alpha$. Therefore, the incidence matrix of an FR code has constant row sum $\alpha$, and constant column sum $\rho$. Notice that a fractional repetition code is also known as a \textit{tactical configuration} in the literature of combinatorial designs. Let $n$ denote the number of blocks, and $v$ be the number of points in an FR code. We refer to such an FR code as an $(n,\alpha, v ,\rho)$-FR code. These code parameters satisfy the following relationship
\begin{equation}
n\alpha = v\rho.
\end{equation}

\textit{Example 1.} When every block is incident with $\alpha=2$ points, then the notion of an FR code coincides with a $\rho$-regular graph (possibly with multiple edges).

\smallskip

An $(n,\alpha, v ,\rho)$-FR code $C$ is used to distribute data packets across a storage system. We encode a data object into $v$ coded packets by using a proper MDS code, which are then replicated and spread across $n$ storage nodes according to the incidence structure specified by $C$. Each coded packet is associated with a certain point in $\cP$, and the storage nodes are associated with the blocks in $\cB$. A packet is stored in a node if and only if~the corresponding point and block are incident. Hence, each node contains $\alpha$ packets, and each packet is stored in $\rho$ nodes. Fig.~\ref{Encoding} shows the encoding process of $C$. Since a packet is replicated $\rho$ times, the system can tolerate any $\rho-1$ node failures.

\smallskip

The \textit{dual} of $C$ is defined as the FR code $(\cB,\cP,\cI^t)$, where  $\cI^t$ is the subset of $\cB\times \cP$ defined by
$$
\cI^t := \{(B,p):\, (p,B)\in \cI\}.
$$
Let $C^t$ denote the dual of $C$. We note that the incidence matrix of $C$ and $C^t$ are the transpose of each other. In \cite{key-6}, the authors refer to the dual FR code as the \textit{Transpose code}.

\begin{lem}
Let $C$ be an $(n,\alpha, v, \rho)$-FR code.

(i) The dual code of $C$ is a $(v, \rho, n, \alpha)$-FR code.

(ii) The double dual of $C$ is $C$ itself.
\label{lemma:easy}
\end{lem}

\textit{Example 2.} Let $C$ be the incidence structure obtained from the line graph of the complete graph on five vertices. This~gives the $(5,4,10,2)$-FR code with incidence matrix
$$
{\begin{bmatrix}
1 & 1 & 1 & 1 & 0 & 0 & 0 & 0 & 0 & 0\\
1 & 0 & 0 & 0 & 1 & 1 & 1 & 0 & 0 & 0\\
0 & 1 & 0 & 0 & 1 & 0 & 0 & 1 & 1 & 0\\
0 & 0 & 1 & 0 & 0 & 1 & 0 & 1 & 0 & 1\\
0 & 0 & 0 & 1 & 0 & 0 & 1 & 0 & 1 & 1
\end{bmatrix}
}
$$
as discussed in \cite{key-5}. We note that this is a $5\times 10$ matrix with constant row sum $\alpha=4$ and constant column sum $\rho=2$. The dual of this FR code is a $(10,2,5,4)$-FR code with incidence matrix
$$
{
\begin{bmatrix}
1 & 1 & 0 & 0 & 0\\
1 & 0 & 1 & 0 & 0\\
1 & 0 & 0 & 1 & 0\\
1 & 0 & 0 & 0 & 1\\
0 & 1 & 1 & 0 & 0\\
0 & 1 & 0 & 1 & 0\\
0 & 1 & 0 & 0 & 1\\
0 & 0 & 1 & 1 & 0\\
0 & 0 & 1 & 0 & 1\\
0 & 0 & 0 & 1 & 1
\end{bmatrix}
}.
$$

\section{The Hierarchy of Supported File Size}

We now formally define the supported file size $M_k(C)$ of~an $(n,\alpha,v,\rho)$-FR code $C = (\cP,\cB,\cI)$. For $k=1,2,\ldots, n$, the \textit{supported file size} $M_k(C)$ of $C$ is defined as
\begin{equation}
M_k(C) := \min_{\mathcal{K}\subset \mathcal{B}, |\mathcal{K}|=k} |\{p\in \cP:\, \exists B \in \mathcal{K}, (p, B) \in \cI \}|,
\label{eq:M}
\end{equation}
where the minimum is taken over all $k$-subsets $\cK$ of the block set $\cB$. The parameter $k$ is called the \textit{reconstruction degree}. By definition, the value of $M_k(C)$ essentially refers to the~smallest number of distinct packets one can download from $k$ storage nodes. For a fixed reconstruction degree $k$, we can encode the data object by an MDS code with parameters $[v,M_k(C)]$, such that the original data can be recovered from any $M_k(C)$ out of the $v$ coded symbols. Treating each symbol as a coded packet, we can distribute the coded packets to $n$ storage nodes~based on the FR code $C$ such that any $k$ storage nodes are sufficiently in decoding the data object.

If we increase the value of $k$, then the resulting supported file size also increases, i.e.,
\begin{equation}
\alpha = M_1(C) \leq M_2(C) \leq \cdots \leq M_n(C)=v.
\label{eq:chain1}
\end{equation}

We call the above the \textit{hierarchy of supported file size} of $C$. We shall also define $M_0(C):=0$ by convention. Similarly,~the file size hierarchy of the dual code $C^t$ is
\begin{equation}
\rho = M_1(C^t) \leq M_2(C^t) \leq \cdots \leq M_v(C^t)=n.
\label{eq:chain2}
\end{equation}

Notice that there is a close relationship between $M_k(C)$ and $M_\ell(C^t)$. This property can be seen from the fact that if we~can find an $x\times y$ all-zero submatrix in the incidence matrix of $C$, then we have
$$
M_x(C) \leq  v-y \text{ and } M_y(C^t) \leq n-x.
$$
This motivates us to define
$$
N_k(C) := |\cP| - M_k(C)
$$
\begin{equation}
=\max_{\cK \subset \cB, |\cK|=k} |\{p\in \cP: \not\exists B\in \mathcal{K}, (p,B)\in \mathcal{I}\}|
\label{NkC}
\end{equation}
with the maximum taken over all subsets $\mathcal{K}\subset \cB$ of size $k$.~By definition, $N_k(C)$ is the largest integer $\ell$ such that we can find an $k \times \ell$ all-zero submatrix in the incidence matrix of $C$. From \eqref{eq:chain1} and \eqref{eq:chain2}, we have
\begin{gather*}
v = N_0(C) > N_1(C) \geq N_2(C) \geq \cdots \geq N_n(C) = 0, \\
n = N_0(C^t) > N_1(C^t) \geq N_2(C^t) \geq \cdots \geq N_v(C^t) = 0 .
\end{gather*}

\begin{figure}
\centering{}\includegraphics[scale=0.105]{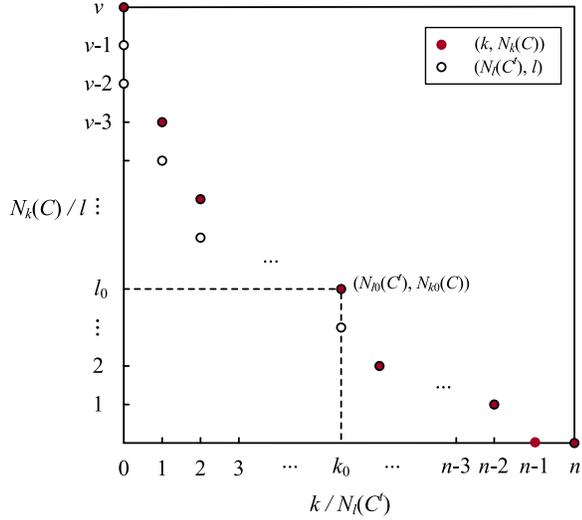}
\caption{The graphic illustration of $(k,N_k(C))$ and $(N_\ell(C^t),\ell)$.}
\label{Pareto}
\end{figure}

We shall plot the points $(k,N_k(C))$ for $k=0,1,\ldots, n$, and $(N_\ell(C^t),\ell)$ for $\ell=0,1,\ldots, v$ in the same figure. The results can be found in Fig.~\ref{Pareto}. A Pareto optimal point, say $(k_0,\ell_0)$,~is a vertex of the graph that satisfies
$$
 \ell_0 = N_{k_0}(C) \text{ and } k_0 = N_{\ell_0}(C^t),
$$
and
\begin{gather*}
N_k(C) < N_{k_0}(C) \text{ for all } k > k_0, \\
N_\ell(C^t) < N_{\ell_0}(C^t) \text{ for all } \ell > \ell_0.
\end{gather*}
Therefore, we obtain
\begin{equation}
N_k(C) = \begin{cases}
v, & \text{for } k=0, \\
v-1, & \text{for } 0 = N_v(C^t) < k \leq N_{v-1}(C^t), \\
v-2, & \text{for } N_{v-1}(C^t) < k \leq N_{v-2}(C^t), \\
\vdots & \vdots \\
1, & \text{for } N_2(C^t) < k \leq N_{1}(C^t), \\
0, & \text{for } N_1(C^t)< k \leq N_0(C^t) = n.
\end{cases}
\end{equation}

Based on the analysis, we obtain the following theorem.
\begin{thm}
Let $C$ be an $(n,\alpha,v,\rho)$-FR code. With $N_\ell(C^t)$~as defined in \eqref{NkC}, we have
\begin{equation}
M_k(C) = \begin{cases}
v, & \text{for } N_1(C^t) <k \leq n = N_0(C^t), \\
v-1, & \text{for } N_2(C^t) <k \leq  N_1(C^t), \\
v-2, & \text{for } N_3(C^t) <k \leq  N_2(C^t), \\
\vdots & \vdots \\
2, & \text{for } N_{v-1}(C^t) <k \leq  N_{v-2}(C^t),\\
1, & \text{for } N_v(C^t)=0 <k \leq  N_{v-1}(C^t).
\end{cases}
\label{eq:Mk}
\end{equation}
\label{thm:duality}
\end{thm}

\begin{rem}
Note that the identities in \eqref{eq:Mk} can be expressed in~a more compact way by
\begin{equation}
M_k(C) = \sum_{i=1}^v \mathbb{I}( k > N_i(C^t) ),
\label{eq:indicator}
\end{equation}
where $\mathbb{I}(P)$ is the indicator function equal to $1$ if the condition $P$ is true and $0$ otherwise. In this case, the right-hand side term of \eqref{eq:indicator} counts the number of $i\in\{1,2,\ldots, v\}$ such that $N_i(C^t)$ is strictly less than $k$. Thus,
$$
\sum_{i=1}^v \mathbb{I}( k > N_i(C^t)) = v-\ell \text{ for } N_{\ell+1}(C^t) < k \leq N_\ell(C^t),
$$
where $k=1,2,\ldots, n$.
\end{rem}

\smallskip

\textit{Example 2 (continued).} For $k=1,2,\ldots, 5$, we can compute that the supported file size $M_k(C)$ of the complete graph based FR code $C$ is
$$
M_k(C) = \begin{cases}
10, & \text{for }k= 4, 5, \\
9, & \text{for } k = 3, \\
7, & \text{for } k = 2, \\
4, & \text{for } k = 1,
\end{cases}
$$
and the values of $N_\ell(C^t)$ for $\ell=1,2,\ldots,10$ are
\begin{align*}
0& =N_7(C^t) = N_8(C^t)=N_9(C^t)=N_{10}(C^t), \\
1&= N_4(C^t)=N_5(C^t)=N_6(C^t), \\
2& =N_2(C^t) = N_3(C^t), \\
3&=N_1(C^t) .
\end{align*}

\begin{figure}
\centering{}\includegraphics[scale=0.102]{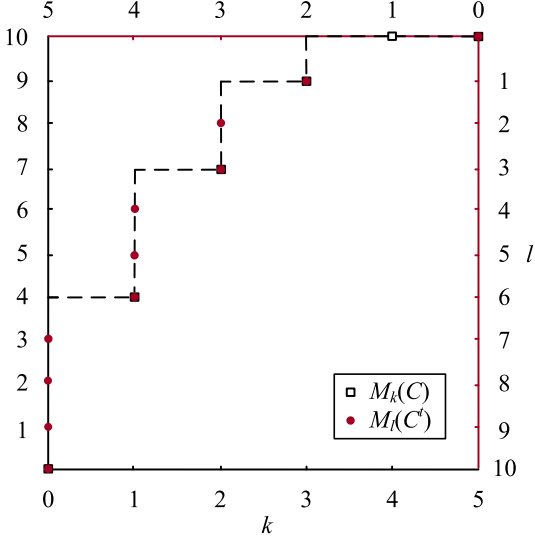}
\caption{The relationship between $M_k(C)$ and $M_\ell(C^t)$.}
\label{Example2}
\end{figure}

Furthermore, the supported file size hierarchy of $C^t$ is $5-N_\ell(C^t)$, i.e,
\begin{align*}
5 &= M_{10}(C^t) = M_{9}(C^t) =M_{8}(C^t) = M_{7}(C^t), \\
4 &= M_{6}(C^t) = M_{5}(C^t) =M_{4}(C^t), \\
3 &= M_{3}(C^t) = M_{2}(C^t),\\
2 &= M_{1}(C^t).
\end{align*}
Fig.~\ref{Example2} illustrates the relationship between $M_k(C)$ and $M_\ell(C^t)$. We can obtain the two supported file size functions if we view the stair-case graph from two different perspectives, which are distinguished with different colors.

\section{Dual Bound on Supported File Size}

Two upper bounds on the supported file size of FR codes are presented in~\cite{key-6}. Specifically, the supported file size $M_k(C)$~of an $(n,\alpha,v,\rho)$-FR code $C$ is upper bounded by
\begin{equation}
M_k(C) \leq \Big\lfloor v \Big(1 - \frac{\binom{n-\rho}{k}}{\binom{n}{k}}\Big) \Big\rfloor,
\label{eq:bound1}
\end{equation}
and
\begin{equation}
M_k(C) \leq g(k),
\label{eq:bound2}
\end{equation}
where $g(k)$ is defined recursively by
$$
g(1) := \alpha, \ g(k+1) := g(k) + \alpha - \Big\lceil \frac{\rho g(k) - k \alpha}{n-k}\Big\rceil.
$$
The bound in \eqref{eq:bound2} is not explicit and is relatively more difficult to compute as compared to \eqref{eq:bound1}. However, it can be shown that  \eqref{eq:bound2} is tighter than \eqref{eq:bound1} in general.

We note that Theorem~\ref{thm:duality} provides a link between an FR code and its dual. Using the mechanism in the previous section, we give an alternate proof of a bound on the reconstruction degree $k$, which is first obtained in~\cite{key-12}.

\begin{thm} (\cite[Lemma 32]{key-12})
\label{thm:Silberstein}
If we store a data object of size $M$ by using an $(n,\alpha,v,\rho)$-FR code $C$, then the reconstruction degree $k$ is lower bounded by
\begin{equation}
k \geq \Big\lceil \frac{n \binom{M-1}{\alpha}}{\binom{v}{\alpha}}\Big\rceil + 1.
\end{equation}
\end{thm}

\begin{IEEEproof}
By applying the bound in \eqref{eq:bound1} to the dual code of $C$, we obtain
\begin{equation}
M_\ell(C^t) \leq  n \Big(1 - \frac{\binom{v-\alpha}{\ell}}{\binom{v}{\ell}}\Big),
\end{equation}
for $\ell=1,2,\ldots, v$. (We can remove the floor operator without loss of generality.) Hence,
\begin{equation}
N_\ell(C^t) \geq n - n \Big(1 - \frac{\binom{v-\alpha}{\ell}}{\binom{v}{\ell}}\Big) = n \frac{\binom{v-\alpha}{\ell}}{\binom{v}{\ell}}.
\end{equation}

Given an integer $M$ between $1$ and $v$, we let $\ell$ be the integer that satisfies
$$
M = v-\ell+1.
$$
By Theorem~\ref{thm:duality}, we obtain
\begin{equation}
k \geq N_\ell(C^t) +1 \geq n \frac{\binom{v-\alpha}{v-M+1}}{\binom{v}{v-M+1}} +1 = n\frac{\binom{M-1}{\alpha}}{\binom{v}{\alpha}} + 1.
\end{equation}
The proof of this theorem is completed by taking the ceiling~of both sides.
\end{IEEEproof}

The above proof illustrates that Theorem~\ref{thm:Silberstein} is essentially the same as the basic bound \eqref{eq:bound1} on the supported file size. We can obtain a slight improvement if we use the bound given in \eqref{eq:bound2} instead.

\begin{thm}
Given an FR code $C$ with parameters $(n,\alpha,v,\rho)$, we define the function $g'(\ell)$ recursively by
$$
g'(1) := \rho,\ g'(\ell+1) := g'(\ell) + \rho - \Big\lceil \frac{\alpha g'(\ell) - \ell \rho}{v-\ell}\Big\rceil,
$$
for $\ell=1,2,\ldots, v-1$.
Then, for all $k = 1,2,\ldots, n$, we have
\begin{equation}
M_k(C) \leq \sum_{i=1}^v \mathbb{I}( k > n-g'(i)).
\label{eq:bound3}
\end{equation}
\end{thm}

\begin{IEEEproof}
We notice that the function $g'(\ell)$ is the counterpart of the recursive bound \eqref{eq:bound2} on the dual code. Thus,
\begin{equation}
M_\ell(C^t) \leq g'(\ell).
\end{equation}
Since
\begin{equation}
N_\ell(C^t) \geq n - g'(\ell),
\end{equation}
for all $\ell$, in view of the remark after the proof of Theorem~\ref{thm:duality}, we have
\begin{equation}
M_k(C) = \sum_{\ell=1}^v \mathbb{I}(k > N_\ell(C^t)) \leq \sum_{\ell=1}^v \mathbb{I}(k > n - g'(\ell)),
\end{equation}
which completes the proof.
\end{IEEEproof}

We refer to the inequality in \eqref{eq:bound3} as the \textit{dual bound} on the supported file size.

\smallskip

\textit{Example 3.} Consider an explicit FR code $C$ with parameters $(n,\alpha,v,\rho) = (15,2,10,3)$. The bounds in \eqref{eq:bound1} and \eqref{eq:bound2} suggest that the supported file size with reconstruction degree $k=6$~is upper bounded by
\begin{align*}
M_6(C) & \leq \Big\lfloor 10\big(1- \binom{12}{6}/\binom{15}{6}\big)\Big\rfloor = 8, \text{ and }\\
M_6(C) & \leq g(6) = 7, \text{ respectively}.
\end{align*}
Moreover, the recursive bound applied to the dual code yields $M_\ell(C^t) \leq g'(\ell)$ with
\begin{gather*}
g'(1)=3,\  g'(2)=5,\ g'(3)=7,\ g'(4)=9,\ g'(5)= 11,\\
g'(6)= 12, \ g'(7)=13,\ g'(8)=14,\   g'(9)=g'(10)=15.
\end{gather*}
Then, the dual bound in \eqref{eq:bound3} gives
$$
M_6(C) \leq \sum_{i=1}^{10} \mathbb{I}(6 > 15-g'(i)) = \sum_{i=1}^{10} \mathbb{I}(g'(i) > 9) = 6.
$$
This bound can be achieved by the $(15,2,10,3)$-FR code listed in the database in \cite{key-10} with the following incidence matrix:

$${\left[\begin{array}{cccccccccc}
1& 1& 0& 0& 0& 0& 0& 0& 0& 0\\
1& 0& 1& 0& 0& 0& 0& 0& 0& 0\\
1& 0& 0& 1& 0& 0& 0& 0& 0& 0\\
0& 1& 0& 0& 1& 0& 0& 0& 0& 0\\
0& 1& 0& 0& 0& 1& 0& 0& 0& 0\\
0& 0& 1& 0& 0& 0& 1& 0& 0& 0\\
0& 0& 1& 0& 0& 0& 0& 1& 0& 0\\
0& 0& 0& 1& 0& 0& 0& 0& 1& 0\\
0& 0& 0& 1& 0& 0& 0& 0& 0& 1\\
0& 0& 0& 0& 1& 1& 0& 0& 0& 0\\
0& 0& 0& 0& 1& 0& 1& 0& 0& 0\\
0& 0& 0& 0& 0& 1& 0& 1& 0& 0\\
0& 0& 0& 0& 0& 0& 1& 0& 1& 0\\
0& 0& 0& 0& 0& 0& 0& 1& 0& 1\\
0& 0& 0& 0& 0& 0& 0& 0& 1& 1
\end{array} \right]
}.
$$

\smallskip

We note that the six storage nodes associated to rows $1,2, 3,$ $ 4, 5$, and $10$ contain precisely $6$ distinct packets. Therefore,~this FR code can support a file size of $M=6$ with $k=6$, implying that it is optimal by the dual bound.

\section{Conclusion}

In this paper, we investigate fractional repetition codes with an emphasize on the duality of FR codes. We start by studying the relationship between the supported file size of an FR code and its dual code. Furthermore, we present an improved upper bound on the supported file size of FR codes, which we refer~to as the dual bound.

Even though we focus on FR codes whose incidence matrix has constant row sum and also constant column sum, the main Theorem~\ref{thm:duality} works for any incidence matrix based code. These codes can be viewed as a generalization of FR codes, and are called \textit{heterogeneous} FR codes. Explicit constructions of such codes can be found in \cite{key-17}\textendash{}\cite{key-21}. The relationship between the file size of a heterogeneous FR code and its dual code can be determined following a similar procedure as in Theorem~\ref{thm:duality}.

\section*{Acknowledgment}

This work was partially supported by the National Natural Science Foundation of China under Grant 61671001, National Key R\&D Program of China under Grant 2016YFB0800101 and 2017YFB0803204, Shenzhen Basic Research Project under Grant 20170306092030521 and 20150331100723974, and a grant from the University Grants Committee of Hong Kong Special Administrative Region (Project No. AoE/E-02/08).

\end{document}